\documentclass[
twocolumn,
 amsmath,amssymb,
 aps,
prl,
floatfix,
]{revtex4-2}

\usepackage{dcolumn}
\usepackage{bm}
\usepackage[paperwidth=210mm,paperheight=297mm,centering,hmargin=2.3cm,tmargin=2.8cm,bmargin=3.4cm]{geometry}
\usepackage{booktabs}
\usepackage{subfigure}
\usepackage{graphicx,epsf}
\usepackage{amsfonts}
\usepackage{amssymb}
\usepackage{amsmath}
\usepackage{overpic}
\usepackage{braket}
\usepackage{xcolor}
\usepackage{cancel}
\usepackage{fancyhdr}
\usepackage[normalem]{ulem}

\makeatletter
\def\maketitle{
\@author@finish
\title@column\titleblock@produce
\suppressfloats[t]}
\makeatother

\begin{document}

\title{Microscopic imaging homogeneous and single phase superfluid density in UTe$_2$}

\author{Yusuke Iguchi$^{1,2}$, Huiyuan Man$^{1,3}$, S. M. Thomas$^4$, Filip Ronning$^4$, Priscila F. S. Rosa$^4$, and Kathryn A. Moler$^{1,2,5}$}

\affiliation{$^1$Geballe Laboratory for Advanced Materials, Stanford University, Stanford, California 94305, USA\\
$^2$Stanford Institute for Materials and Energy Sciences, SLAC National Accelerator Laboratory, 2575 Sand Hill Road, Menlo Park, California 94025, USA\\
$^3$Stanford Nano Shared Facilities, Stanford University, Stanford, CA 94305, USA\\
$^4$Los Alamos National Laboratory, Los Alamos, New Mexico 87545, USA\\
$^5$Department of Applied Physics, Stanford University, Stanford, California 94305, USA}

\begin{abstract}
The spin-triplet superconductor UTe$_2$ shows spontaneous time-reversal symmetry breaking and multiple superconducting phases in some crystals, implying chiral superconductivity. Here we microscopically image the local magnetic fields and magnetic susceptibility near the surface of UTe$_2$, observing a homogeneous superfluid density $n_s$ and homogeneous pinned vortices. The temperature dependence of $n_s$ is consistent with an anisotropic gap and shows no evidence for an additional kink that would be expected at any second phase transition.  Our findings are consistent with a dominant $B_{3u}$ superconducting order parameter in the case of a quasi-2D Fermi surface and provide no evidence for multiple phase transitions in $n_s(T)$ in UTe$_2$.
\end{abstract}

\maketitle



Strong spin-orbit coupled unconventional superconductors, whose superconducting (SC) state cannot be described by electron-phonon coupling, provide a platform for experimental and theoretical studies of emergent quantum behavior \cite{Pfleiderer2009,Smidman2017}. Time-reversal and parity are key symmetries to characterize these materials, and striking states of matter often emerge when one (or both) of these symmetries are broken. For instance, odd-parity superconductors have been identified as a promising route for topological superconductivity, which hosts edge modes or vortices with non-abelian statistics required for topological quantum computing~\cite{SatoAndo2017}. A chiral superconductor further breaks time-reversal symmetry and lowers the energy of the SC condensate by removing nodes from the gap function~\cite{Kallin2016}. Odd-parity chiral superconductors are remarkably rare, but their experimental manifestation has been observed in superfluid $^3$He and actinide superconductor UPt$_3$~\cite{Avers2020Broken}. 

UTe$_2$ is a newly-discovered candidate for odd-parity chiral superconductivity~\cite{Ran2019Nearly,Aoki2022Unconventional}. Nuclear magnetic resonance (NMR) Knight shift measurements strongly suggest that UTe$_2$ is an odd-parity superconductor with a dominant $B_{3u}$ order parameter~\cite{Nakamine2019Sup,Nakamine2021Ani,Fujibayashi2022Super}. Point nodes in the SC gap structure are supported by transport measurements~\cite{Ran2019Nearly,Metz2019Point,Kittaka2020Ori}, Knight shift measurements~\cite{Fujibayashi2022Super}, and non-local superfluid density measurements~\cite{Bae2021Ano}. 
The position of the point nodes, however, is still controversial. Thermal conductivity, microwave surface impedance, and specific heat measurements suggest point nodes in the $ab$-plane or along $a$~\cite{Metz2019Point,Bae2021Ano,Kittaka2020Ori}, whereas magnetic penetration depth measurements argue for a multicomponent SC state with multiple point nodes near the $b$- and $c$-axes~\cite{Ishihara2021Chiral}.  

Evidence for chiral superconductivity was found in UTe$_2$ by scanning tunneling spectroscopy on the step edges of a ($0\bar{1}1$)-plane~\cite{Jiao2020Chiral} and by polar Kerr rotation measurements~\cite{Hayes2021Multi}. Multiple SC phase transitions were also reported in UTe$_2$ even at ambient pressure by specific heat measurements~\cite{Hayes2021Multi, Thomas2020}. Subsequently, it was found that the observed ``double peak" in the specific heat can arise from sample inhomogeneity~\cite{Thomas2021Spat}. In addition, a single phase transition is reported in higher quality samples with higher  SC critical temperature  $T_{\rm c}$, higher residual resistivity ratios, lower residual resistivities, and quantum oscillations~\cite{Rosa2021Single,Aoki2022Unconventional}. 

To microscopically investigate the SC state of UTe$_2$, here we report the temperature dependence of the local superfluid response using scanning SQUID (superconducting quantum interference device) susceptometry on a cleaved (011)-plane of UTe$_2$. We also image the pinned vortices induced by field cooling. Our results show no evidence for multiple phase transitions in the temperature dependence of the superfluid density and imply an anisotropic nodal gap structure in UTe$_2$.

\begin{figure*}[htb]
\begin{center}
\includegraphics*[width=16cm]{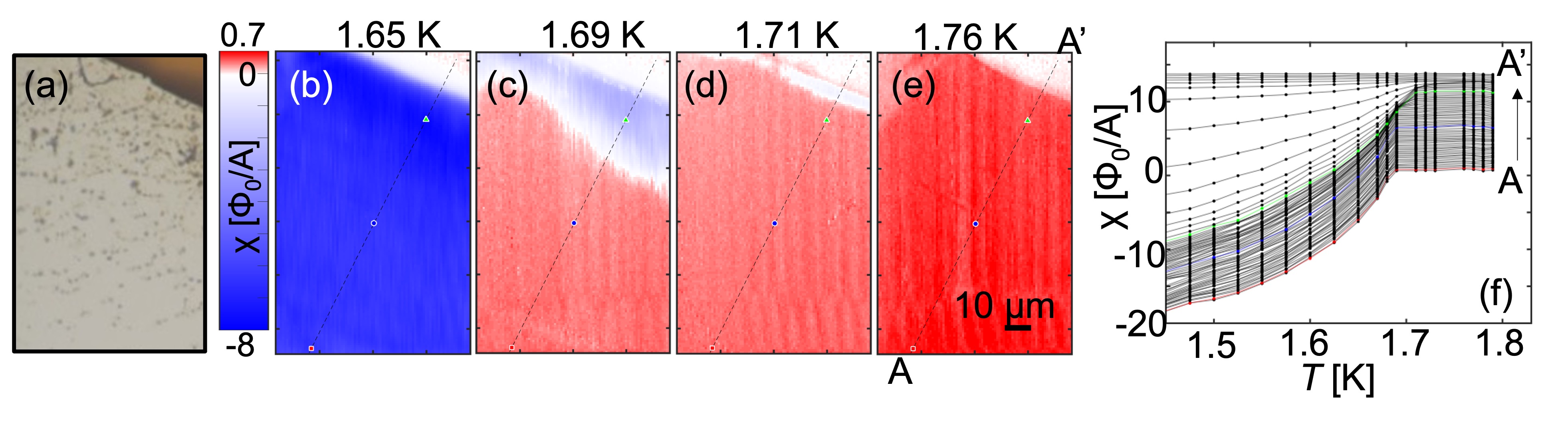}
\caption{ Local susceptibility is microscopically homogeneous on (011) surface of UTe$_2$ sample\#1. (a) Optical image of the scanned area, which includes the cleaved (011) surface with small bumps creating no signals in our scans and the edge. (b-e) Temperature dependence of the susceptometry scan indicates the homogeneous superfluid density on UTe$_2$. Stripes along the scan directions are the scanning noise. (f) The temperature dependence of the local susceptibility at the points from A to A' in Fig. 1(e) has no kink below $T_{\rm c}$. The susceptibilities are shifted by 0.2 $\Phi_0$/A for clarity except for the data at A.}
\end{center}
\end{figure*}

Bulk single crystals of UTe$_2$ were grown by chemical vapor transport. Samples \#1 and \#2 used in this paper were obtained from the same batch as sample s2 in Ref.~\cite{Rosa2021Single}. Heat capacity measurements confirmed a single SC transition at $T_{\rm c} =1.68$~K with a width of 50 mK on a single crystal which was subsequently cleaved into two samples used in this study. We used a scanning SQUID susceptometer to obtain the local ac susceptibility on a cleaved (011)-plane of UTe$_2$ at temperatures from 80 mK to 2 K in a Bluefors LD dilution refrigerator. Our scanning SQUID susceptometer has two pickup loop and field coil pairs configured with a gradiometric structure~\cite{Kirtleyrsi2016}. The inner radius of the pickup loop is 0.4 $\mu$m and the inner radius of the field coil is 1.5 $\mu$m. The scan height is $\sim$500 nm. The pickup loop provides the local dc magnetic flux $\Phi$ in units of the flux quantum $\Phi_0=h/2e$, where $h$ is the Planck constant and $e$ is the elementary charge. The pickup loop also detects the ac magnetic flux $\Phi^{ac}$ in response to an ac magnetic field $He^{i\omega t}$, which was produced by an ac current of $|I^{ac}| =$ 1 mA at a frequency $\omega/2\pi\sim$ 1 kHz through the field coil, using an SR830 Lock-in-Amplifier. Here we report the local ac susceptibility as $\chi=\Phi^{ac}/|I^{ac}|$ in units of $\Phi_0$/A.

To obtain the homogeneity of the superfluid density and its local temperature dependence, we measured the local susceptibility near the edge of the sample [Figs. 1(a)-1(e)]. The susceptibility far from the edge has a homogeneous temperature dependence on the micron-scale [Fig. 1(f)]. We note that our results do not rule out possible inhomogeneity either on the nanoscale or on scales larger than the scan area (e.g  sub-millimeter). Our data also cannot rule out fluctuations in time. There is no kink in the temperature dependence of the local susceptibility below $T_{\rm c}$, wherein the temperature step of the scans is 25 mK. The susceptibility is positive above $T_{\rm c}$ due to paramagnetism. The local susceptibility was negative (diamagnetic) near the edge but positive far from the edge at 1.69 K [Figs. 1(c),1(f)].

We defined the local $T_{\rm c}$ as the temperature that satisfies the relation of $\chi(T>T_{\rm c})>$-0.1 $\Phi_0/$A. The local $T_{\rm c}$ mapping clearly shows that the local $T_{\rm c}$ is weakly enhanced at the edge but is homogeneous 30~$\mu$m away from the edge into the sample [Fig. 2, sample\#1; Fig. S2, sample\#2]. We note that the reported local $T_{\rm c}$ near the edge is a lower bound relative to the actual value because the penetration depth near $T_{\rm c}$ is longer than the pickup loop's scale and the susceptibility loses some signal at the edge. If two phase transitions do exist, they must be closer to each other than 25 mK, or the second kink below $T_{\rm c}$ is much smaller than our experimental noise.

\begin{figure}[!hb]
\begin{center}
\includegraphics*[width=7cm]{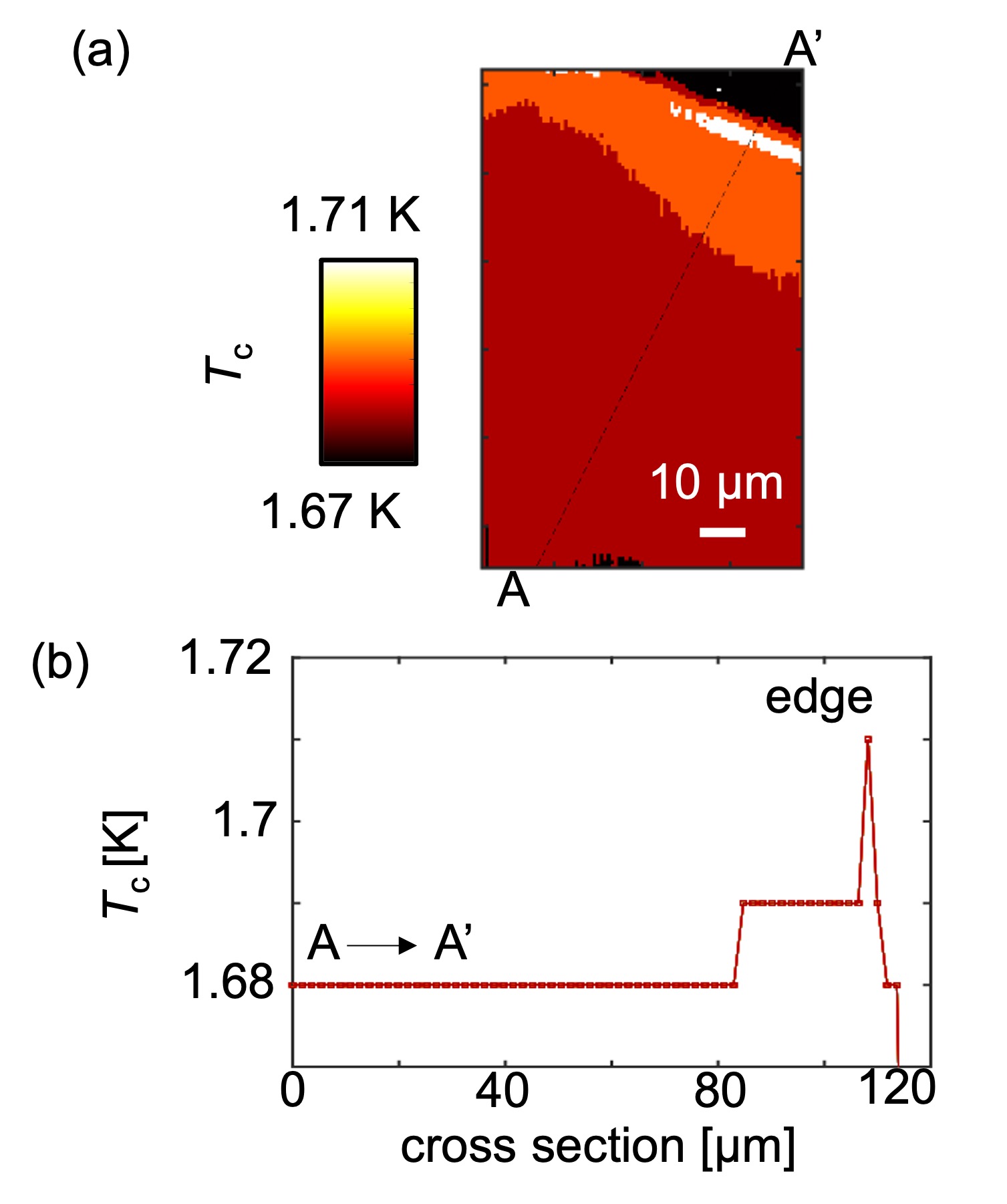}
\caption{ Small enhancement of local $T_{\rm c}$ near the edge of sample\#1. (a) The local $T_{\rm c}$ mapping is obtained from the local susceptometry scans. (b) Cross section of the local $T_{\rm c}$ from A to A' shows the local $T_{\rm c}$ enhancement of 30 mK at the edge. The plotted area and the cross section from A to A' are the same with Fig. 1.}
\end{center}
\end{figure}

\begin{figure*}[htb]
\begin{center}
\includegraphics*[width=13cm]{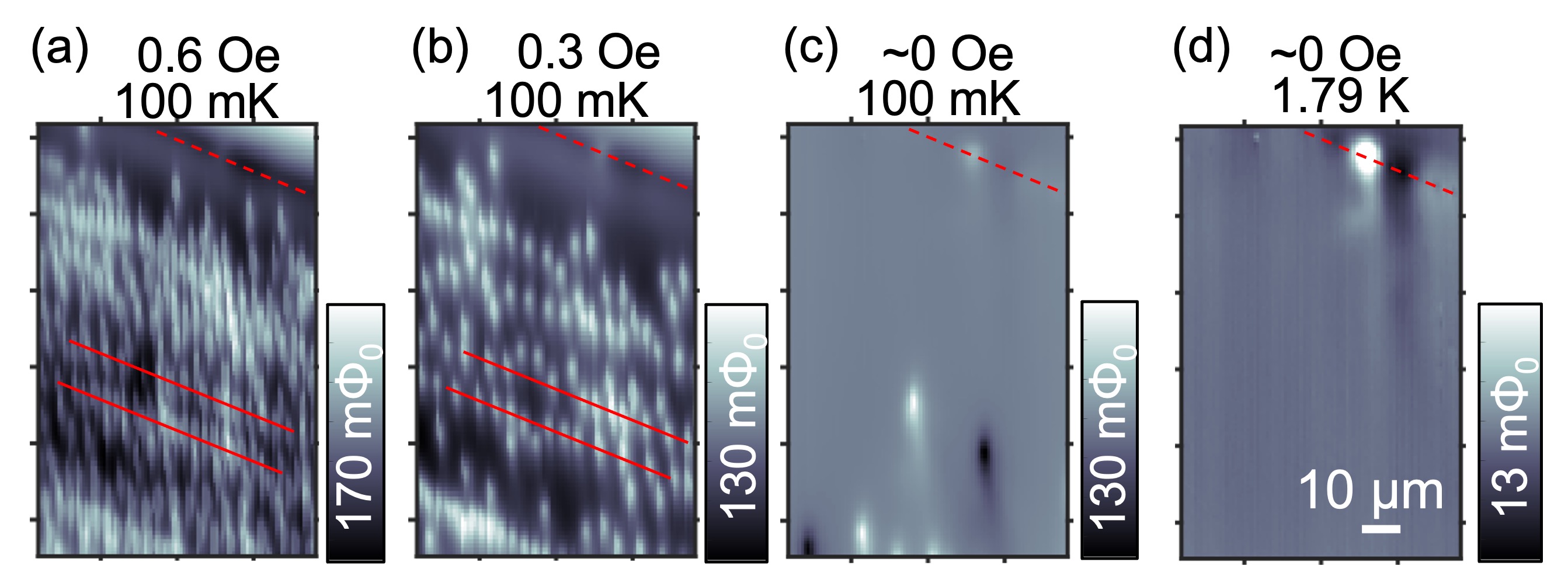}
\caption{The vortex density is homogeneous over many-micron distances. The existence of vortices and antivortices in low-field scans may indicate a local magnetic source in the sample\#1. (a,b) Local magnetometry scan after field cooling shows the vortices pinned parallel to the sample edge, as denoted by the dashed lines. (c) There are a few vortices and antivortices pinned far from the edge after near zero field cooling. (d) Magnetometry scan near zero field above $T_{\rm c}$ shows a small magnetic dipole at the sample's edge, but no other indication of magnetism. The "tail" of the vortices and dipoles are due to the asymmetric shielding structure of the scanning SQUID~\cite{Kirtleyrsi2016}.}
\end{center}
\end{figure*}

Now we turn to the pinned vortex density, which reflects the impurity density on the crystal surface for small applied magnetic fields. The distance between vortices is on the order of microns. Our magnetometry scan imaged the pinned vortices induced by cooling in an applied uniform magnetic field from 2 K to 100 mK [Figs. 3(a),3(b), sample\#1; Figs. S3(a),S3(b), sample\#2]. The number of vortices corresponds to the applied field, but the vortices are preferentially pinned along lines in one direction, which is parallel to the sample edge. This linear pinning indicates the existence of a line anomaly, such as nanometer-scale step edges along crystal axes. Near zero magnetic field, there are still a few vortices and antivortices pinned far from the edge [Fig. 3(c), sample\#1; Fig. S3(c), sample\#2]. Notably, these vortices and antivortices do not disappear after zero field cooling with slower cooling rates, which is expected to cancel the uniform background field normal to the sample surface by the application of an external field. These data are inconsistent with the argument from polar Kerr effect measurements that there are no vortices in UTe$_2$ within the beam size area ($\sim11~\mu$m radius)~\cite{Wei2022}. Further, our results indicate the existence of a local magnetic source that induces vortices and antivortices, in spite of the absence of long-range order or strong magnetic sources on the scan plane above $T_{\rm c}$ \cite{Ran2019Nearly,Miyake2019Meta,Hutanu2020Low}. 
Small dipole fields are observed at the edge of the sample, which may stem from U impurities; however, these impurities cannot induce pinned vortices and antivortices as they are too far away [Fig. 3(d), sample\#1]. Muon spin resonance and NMR measurements have detected the presence of strong and slow magnetic fluctuations in UTe$_2$ at low temperatures~\cite{Tokunaga2022Slow,Sundar2022Ubi}. Therefore, a sensible scenario is that these fluctuations are pinned by defects and become locally static.

\begin{figure*}[htb]
\begin{center}
\includegraphics*[width=16cm]{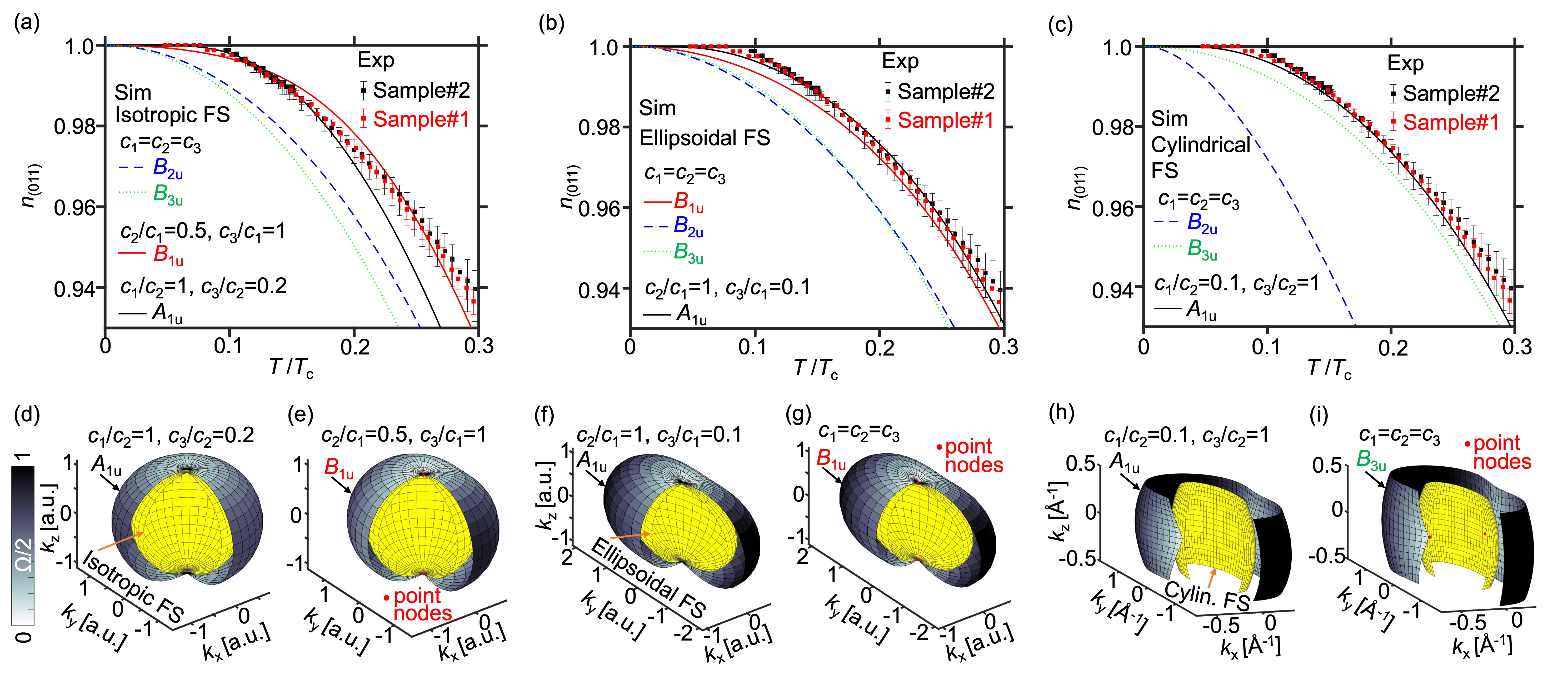}
\caption{The temperature dependence of the  superfluid density = best matches an anisotropic, rather than isotropic, gap structure. (a-c) Temperature dependence of the normalized superfluid density $n_{(011)}$ at the fixed position in sample\#1,\#2 that are indicated by the blue dot in Figs. 1(b), S1(b), respectively. The thick lines are simulation curves best fit for 4 gap symmetries with (a) Isotropic FS, (b) Ellipsoidal FS, and (c) Cylindrical FS~\cite{supple}. (d-i) Best fit models of gap symmetries (d,e), (f,g) and (h,i) for (a), (b) and (c), respectively. FS is plotted by yellow color. The distance between larger surfaces and FS represents the angular dependence of the SC gap $\Omega$ in (a,b) the spherical coordinate and (c) the cylindrical coordinate. All surfaces are cut for clarity.}
\end{center}
\end{figure*}

To estimate the local superfluid density, we measure the local susceptibility at different temperatures with the pickup loop position fixed. The local superfluid density is obtained using the numerical expression of the susceptibility assuming a homogeneous penetration depth, $\lambda$, as described below. Kirtley {\it et al}. developed the expression for the susceptibility as a function of the distance between the susceptometer and the sample surface~\cite{Kirtley2012}. In this model, wherein the sample surface is at $z=0$, we consider three regions.  Above the sample ($z>0$), the pickup loop and field coil are at $z$ in vacuum and $\mu_1 = \mu_0$, where $\mu_0$ is the permeability in vacuum. In the sample ($-t\leq z \leq 0$), the London penetration depth is $\lambda=\sqrt{m/4\pi ne^2}$, and the permeability is $\mu_2$. Below the sample ($z<-t$), there is a nonsuperconducting substrate with a permeability $\mu_3$. The radius of the field coil and the pickup loop are $a$ and $b$, respectively. By solving Maxwell's equations and the London equation for the three regions, the SQUID height dependence of the susceptibility $\chi(z)$ is expressed as 
\begin{widetext}
\begin{equation}\label{phi-z}
    \chi(z)/\phi_s = \int_0^\infty{dx e^{-2x\bar{z}}xJ_1(x)}\left[\frac{-(\bar{q}+\bar{\mu}_2x)(\bar{\mu}_3\bar{q}-\bar{\mu}_2x)+e^{2\bar{q}\bar{t}}(\bar{q}-\bar{\mu}_2x)(\bar{\mu}_3q+\bar{\mu}_2x)}{-(\bar{q}-\bar{\mu}_2x)(\bar{\mu}_3\bar{q}-\bar{\mu}_2x)+e^{2\bar{q}\bar{t}}(\bar{q}+\bar{\mu}_2x)(\bar{\mu}_3q+\bar{\mu}_2x)}\right],
\end{equation}
\end{widetext}
where $\phi_s = A\mu_0/2\Phi_0a$ is the self inductance between the field coil and the pickup loop, $A$ is the effective area of the pickup loop, $\bar{z} = z/a$, $J_1$ is the Bessel function of first order, $\bar{t} = t/a$, $\bar{q} = \sqrt{x^2 + \bar{\lambda}^{-2}}$, and $\bar{\lambda} = \lambda/a$. For the bulk sample on a copper substrate ($\bar{t}>>1, \mu_3 = 1$), the observed susceptibility only depends on $\lambda$, $\mu_2$, and the SQUID structure. 

The penetration depth $\lambda(T)$ was calculated using Eq.~\eqref{phi-z} and the observed susceptibility [Fig. 4(a)]. The normalized superfluid density $n_{s} = \lambda^2(0)/\lambda^2(T)$ was calculated from the obtained penetration depth's temperature dependence, where $\lambda(0) = 1620\pm150$~nm [sample\#1], $1730\pm300$~nm [sample\#2] [Fig.~4(b)]. Here the error for $\lambda$ and $n_s$ is roughly calculated from the pickup loop height uncertainty. We note that sample\#2 had a dead layer of 700~nm on the surface, which we estimated by assuming that sample\#2 has a similar penetration depth at zero temperature with sample\#1, because sample\#2 was accidentally exposed in air about one extra hour. The locally obtained superfluid density $n_s$ saturates below $T/T_{\rm c}=0.1$.

We examine the SC gap structure through the temperature dependence of the superfluid density. The superfluid density $n_{i}$ is sensitive to low-energy excitations along the $i$ axis, which is perpendicular to the applied field. In our case, $n_{i}$ is sensitive to excitations within the plane normal to [011], and the extracted $n_{(011)}$ is the average of $n_a$ and $n_{\perp[011],a}$)~\cite{Chandrasekhar1993}. The SC gap function of UTe$_2$, $\Delta$, is most likely odd-parity within the orthorhombic $D_{2h}$ point group. In the presence of strong spin-orbit coupling, $\Delta(T,\Vec{k})=\Psi(T)\Omega(\Vec{k})$, and the angle dependence of the gap function is expressed as $\Omega(\Vec{k})\propto\sqrt{\Vec{d}\cdot\Vec{d}^*\pm|\Vec{d}\times\Vec{d}^*|}$. In this case, the possible irreducible representations are $A_{1u}$ [full gap, $\Vec{d}=(c_1k_x,c_2k_y,c_3k_z)$)], $B_{1u}$ [point nodes along $c$, $\Vec{d}=(c_1k_y,c_2k_x,c_3k_xk_yk_z)$], $B_{2u}$ [point nodes along $b$, $\Vec{d}=(c_1k_z,c_2k_xk_yk_z,c_3k_x)$], and $B_{3u}$ [point nodes along $a$, $\Vec{d}=(c_1k_xk_yk_z,c_2k_z,c_3k_y)$]~\cite{Annett1990}. We note that coefficients $c_1, c_2$, and $c_3$ may differ by orders of magnitude~\cite{IshizukaPRB2021}. 

For the sake of completeness, here we assume three cases of Fermi surface structure to calculate the temperature-dependent superfluid density with fit parameters $c_1, c_2$, and $c_3$~\cite{supple,Kogan2021}: (I) isotropic Fermi Surface (FS) based on the isotropic heavy 3D Fermi surface observed by angle-resolved photoemission spectroscopy (ARPES) measurements ~\cite{FujimoriJPSJ2019,MiaoPRL2020}; (II) ellipsoidal FS based on the  upper critical field~\cite{AokiJPSJ2019}; and (III) cylindrical FS. Case (III) is based on both ARPES measurements, which observed cylindrical light electron bands~\cite{FujimoriJPSJ2019,MiaoPRL2020} and recent de Haas–van Alphen measurements that reveal heavy cylindrical bands ~\cite{Aoki2022First}. 

The isotropic fully gapped model $A_{1u}$ saturates at $T/T_{\rm c}=0.2$ [Fig.~S6]. In contrast, the experimental data saturate at a lower temperature, which implies an anisotropic structure in the SC gap function. The calculated normalized superfluid density $n_{(011)}\sim(n_a+n_{\perp[011],a})/2$ for highly anisotropic $A_{1u}$ and $B_{1u}$ have a similar temperature dependence compared to our experimental results, whereas $n_{(011)}$ for $B_{2u}$ and $B_{3u}$ do not agree with our data because of their point nodes near the (011) plane with isotropic or ellipsoidal 3D Fermi surfaces [Figs.~4(a),4(b)]. For highly anisotropic $A_{1u}$ and $B_{3u}$, $n_{(011)}$ agrees with our experimental results, whereas $n_{(011)}$ for $B_{2u}$ inconsistent with the data for a cylindrical Fermi surface [Fig.~4(c)]. We note that our calculations with point nodes do not completely explain our results near zero temperature, which may be caused by our assumptions of the simplest structures of Fermi surface and gap functions, the simple averaging of $n_a$ and $n_{\perp[011],a}$, or by the assumption of a single band.
Our results indicate the existence of point nodes along the $a$ axis for a cylindrical Fermi surface. A highly anisotropic fully-gapped component is also allowed.

In summary, we microscopically imaged the superfluid density and the vortex density in high quality samples of UTe$_2$. The superfluid density is homogeneous, and the temperature dependence below the SC transition $T_{\rm c}$ does not show evidence for a second phase transition. The observed temperature dependence of the superfluid density can be explained by a $B_{1u}$ order parameter for a 3D ellipsoidal (or isotropic) Fermi surfaces or by a $B_{3u}$ order parameter for a quasi-2D cylindrical Fermi surface. A highly anisotropic $A_{1u}$ symmetry component is also allowed for any Fermi surface structures. Combining our results with previous studies about the gap symmetry of UTe$_2$~\cite{Bae2021Ano,Metz2019Point,Kittaka2020Ori,Fujibayashi2022Super}, we conclude that the SC order parameter is most likely dominated by the $B_{3u}$ symmetry. In light of our results, evidence for time-reversal symmetry breaking and chiral superconductivity in UTe$_2$ could be understood either through the presence of vortices and antivortices even at zero applied field or by the presence of a finite anisotropic $A_{1u}$ symmetry in the SC order parameter.

\begin{acknowledgments}
The authors thank J. Ishizuka for fruitful discussions. This work was primarily supported by the Department of Energy, Office of Science, Basic Energy Sciences, Materials Sciences and Engineering Division, under Contract No. DE- AC02-76SF00515. Sample synthesis at LANL was supported by the U.S. Department of Energy, Office of Basic Energy Sciences, Division of Materials Science and Engineering “Quantum Fluctuations in Narrow-Band Systems” project, while heat capacity measurements were performed with support from the LANL LDRD program. Y.I. was supported by the Japan Society for the Promotion of Science (JSPS), Overseas Research Fellowship.
\end{acknowledgments}



\section*{Contributions}
Y.I. carried out the scanning SQUID microscopy, analyzed experimental data, simulated the superfluid density, and wrote the manuscript. H.M. carried out the scanning SQUID microscopy. S.M.T., F.R., and P.F.S.R. synthesized the crystals. K.A.M. supervised the project. All the authors discussed the results and implications and commented on the manuscript.
	
	
\section*{Additional information}
Correspondence and requests for materials should be addressed to Y. I. (yiguchi@stanford.edu)



\newpage
\clearpage

\setcounter{figure}{0}
\setcounter{equation}{0}
\renewcommand{\thefigure}{S\arabic{figure}}
\renewcommand{\theequation}{S\arabic{equation}}

\onecolumngrid
\appendix
	
\begin{center}
	\Large
	{Supplemental Material for \\\lq\lq Microscopic imaging homogeneous and single phase superfluid density in UTe$_2$ \rq\rq} \\by Iguchi $et$ $al.$
\end{center}


\begin{figure*}[htb]
\begin{center}
\includegraphics*[width=16cm]{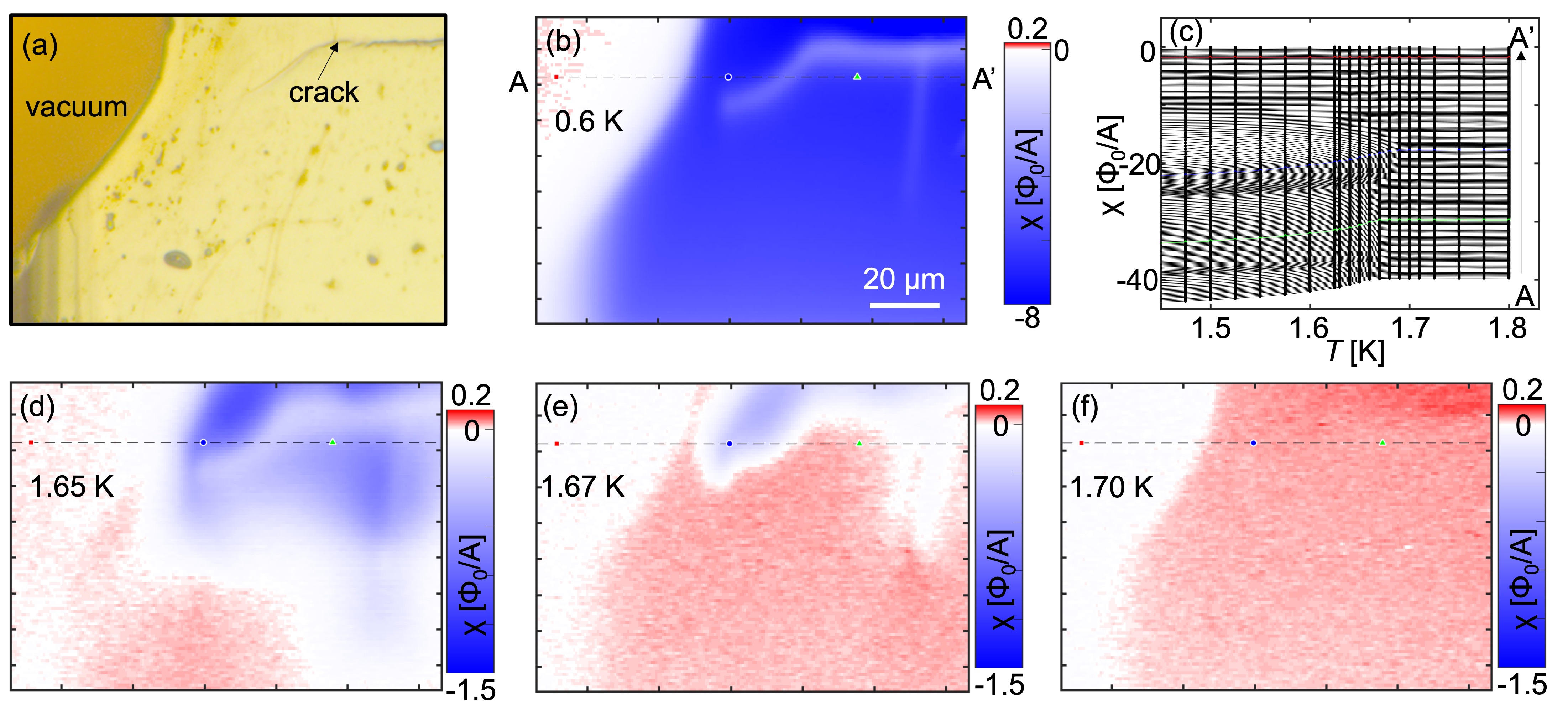}
\caption{Local susceptibility is microscopically homogeneous on (011) surface of UTe$_2$ sample\#2. (a) Optical image of the scanned area, which includes the cleaved (011) surface and the edge. (b) Susceptometry scan at 0.6 K showing homogeneity. (c) The temperature dependence of the local susceptibility at the points from A to A' in Fig. S1(b) has no kink below $T_{\rm c}$. The susceptibilities are shifted by 0.2 $\Phi_0$/A for clarity except for the data at A'. (d-e) Temperature dependence of the susceptometry scan indicates the homogeneous superfluid density and weak enhance of $T_{\rm c}$ near the edge and cracks on UTe$_2$ sample\#2. }
\end{center}
\end{figure*}

\begin{figure*}[htb]
\begin{center}
\includegraphics*[width=14cm]{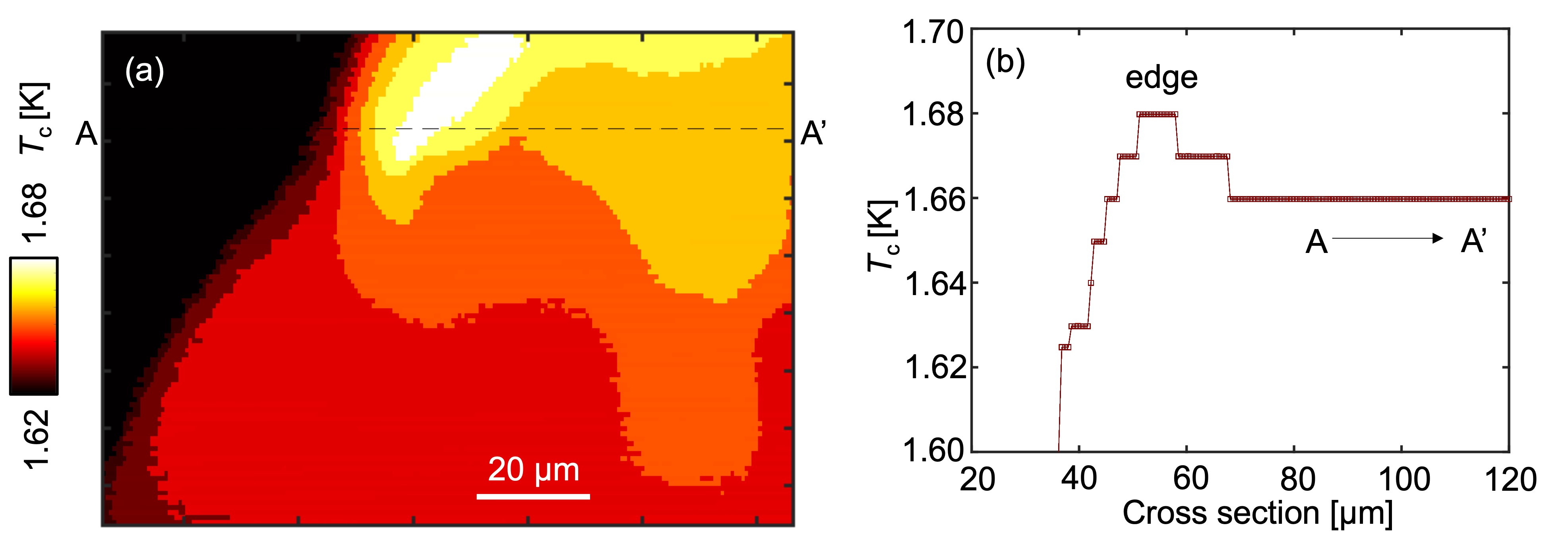}
\caption{Local $T_{\rm c}$ with small enhancement near the edge and cracks in sample\#2. (a) The local $T_{\rm c}$ mapping is obtained from the local susceptometry scans in sample\#2. (b) Cross section of the local $T_{\rm c}$ from A to A' shows the local $T_{\rm c}$ enhancement of 20 mK near the edge.  }
\end{center}
\end{figure*}

\begin{figure*}[htb]
\begin{center}
\includegraphics*[width=14cm]{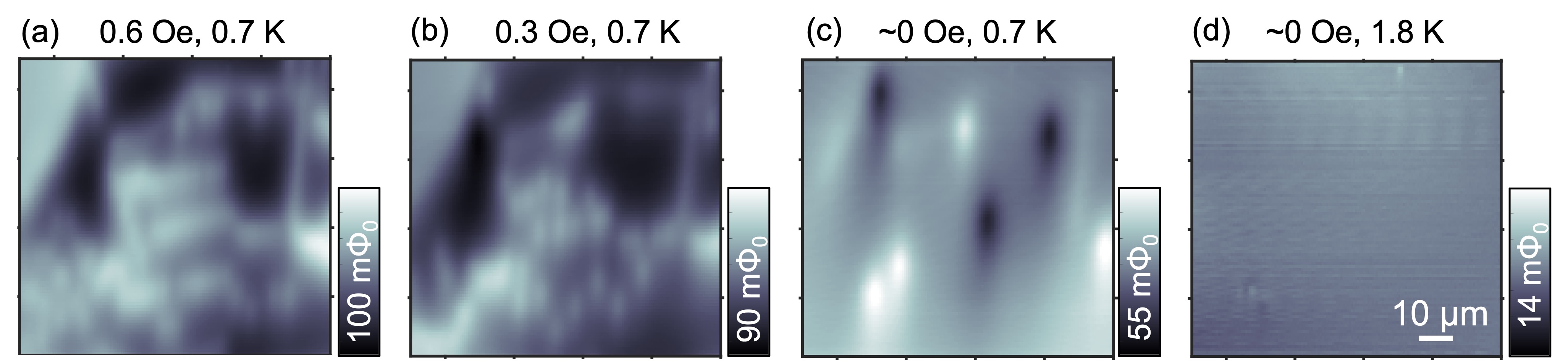}
\caption{Vortex density is homogeneous but the existence of vortex and antivortex indicates the existence of local magnetic source in sample\#2. (a,b) Local magnetometry scan after field cooling shows the some vortices are pinned linearly but others not. This is different from sample\#1, which the local strain due to the cracks may explain. (c) There are vortices and antivortices pinned after near zero field cooling. (d) The magnetometry scan at near zero field above $T_{\rm c}$ shows no strong local magnetic fields.  }
\end{center}
\end{figure*}

\begin{figure*}[tb]
\begin{center}
\includegraphics*[width=8cm]{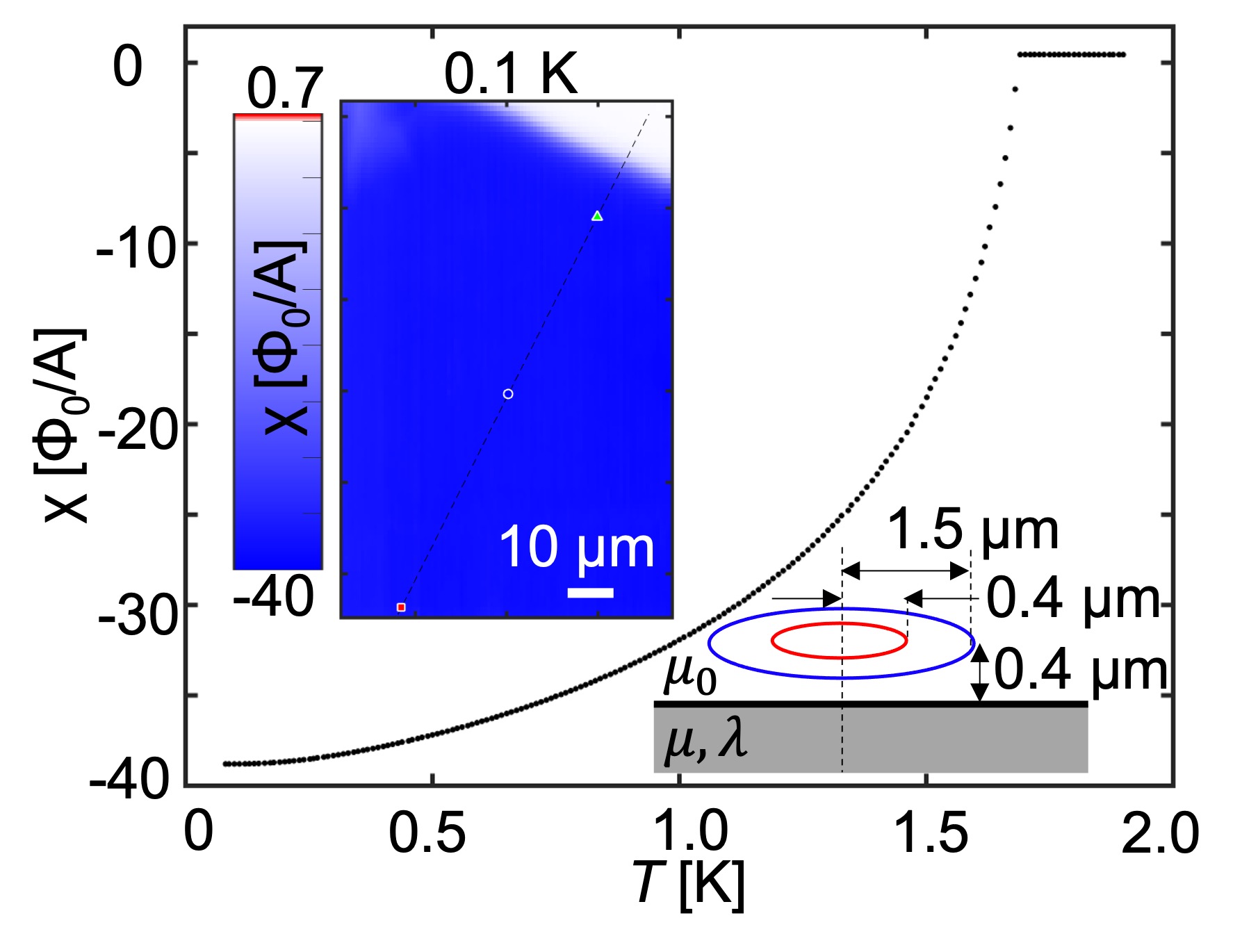}
\caption{Temperature dependence of measured susceptibility at the fixed position in sample\#1 indicated by the blue dot in the inset of susceptibility scan at 100 mK. This data is fitted to obtain the superfluid density in Fig. 4. Inset sketches the experimental set up of the pickup loop, the field coil, and the scan height for the fitting. The self inductance $\phi_s$ is 270 $\Phi_0$/A.}\label{susc-temp}
\end{center}
\end{figure*}

\clearpage

\section{Semi-classical model of generalized London equation in semi-infinite crystal}
\begin{figure*}[tb]
\begin{center}
\includegraphics*[width=8cm]{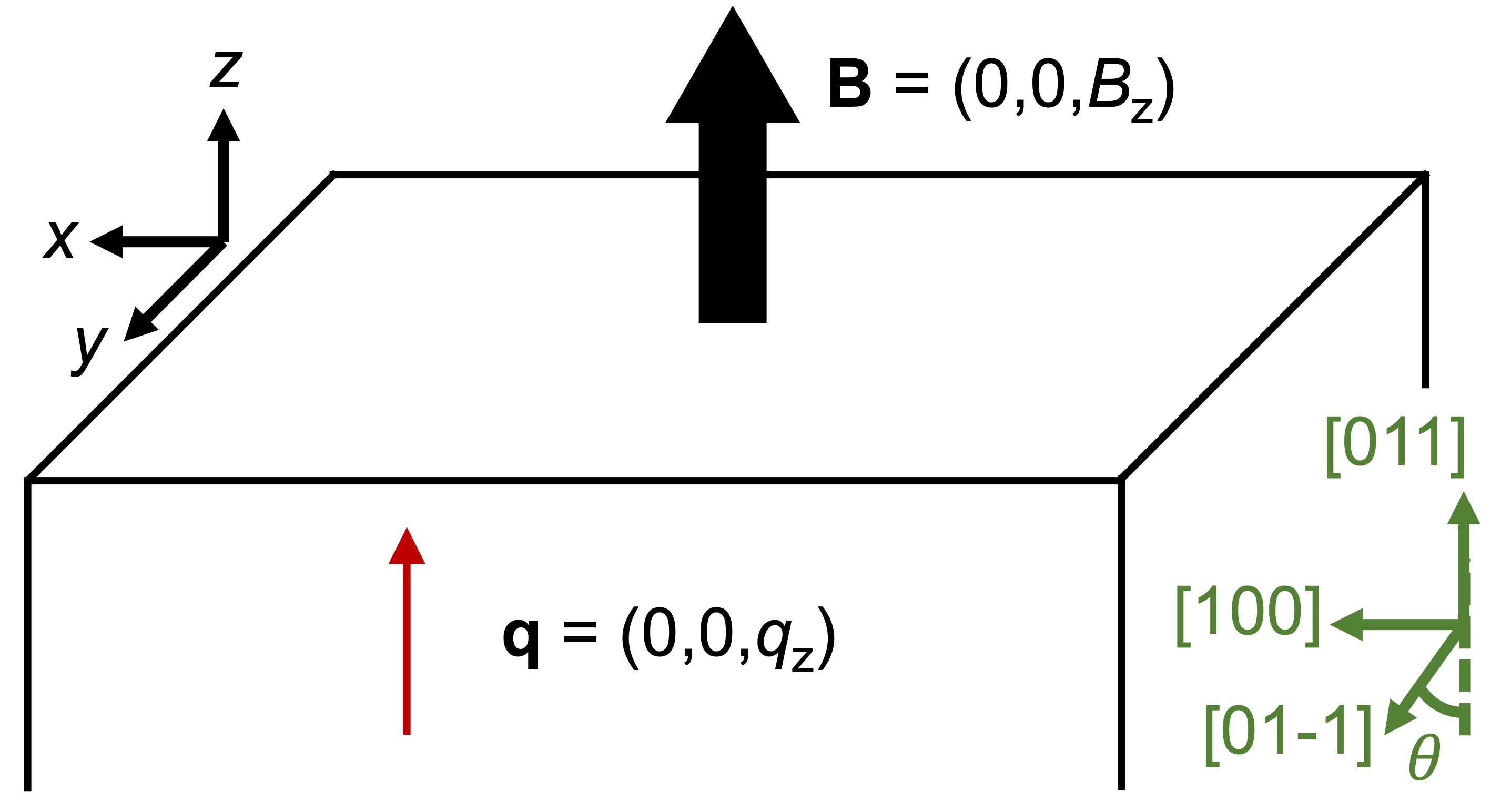}
\caption{Experimental configuration for the superfluid density calculation.}\label{setup}
\end{center}
\end{figure*}

The generalized London equation is expressed by the semi-classical approach~\cite{Chandrasekhar1993} as 
\begin{eqnarray}
    \Vec{j}_s &=& -\frac{e^2}{4\pi^3}\int{d^3k\left( -\frac{\partial n_k}{\partial\varepsilon_k}+\frac{\partial f(E_k)}{\partial E_k} \right)(\Vec{v}_k\Vec{v}_k)\cdot\Vec{A}}\\
    &=& - \tilde{T}\cdot\Vec{A},\label{London1}
\end{eqnarray}
where $\Vec{j}_s$ is the supercurrent, $e$ is the elementary charge, $\Vec{k}$ is a wave vector of electron, $n_k$ is the occupancy of the single-particle state $\Vec{k}$ in the superconducting state, $\varepsilon_k$ is the single-particle energy in state $\Vec{k}$, $f(E_k)$ is Fermi distribution function and shows the quasi-particle occupancy of the state $\Vec{k}$, $E_k=\sqrt{\xi^2_k+\Delta^2_k}$ is the energy of quasi-particle in state $\Vec{k}$, $\xi_k=\varepsilon_k-\mu$ is the single-particle energy from Fermi energy, $\mu$ is the chemical potential, $\Delta_k$ is the superconducting gap energy at state $\Vec{k}$, $\Vec{v}_k$ is the electron density at state $\Vec{k}$, and $\Vec{A}$ is the vector potential. $\tilde{T}$ is a symmetric tensor. The tensor $\Vec{v}_k\Vec{v}_k$ is defined as
\begin{equation}
    \Vec{v}_k\Vec{v}_k = 
     \begin{pmatrix}
  v^2_{kx} & v_{kx}v_{ky} & v_{kx}v_{kz} \\
  v_{kx}v_{ky} & v^2_{ky} & v_{ky}v_{kz} \\
  v_{kx}v_{kz} & v_{ky}v_{kz} & v^2_{kz}
 \end{pmatrix}.
\end{equation}
For a semi-infinite superconductor, the complete form of Eq.~(\ref{London1}) satisfying charge conservation is expressed as
\begin{equation}
    \Vec{j}_s = -\left( \tilde{T} - \frac{(\tilde{T}\cdot\Vec{q})(\Vec{q}\cdot\tilde{T})}{\Vec{q}\cdot\tilde{T}\cdot\Vec{q}} \right) \cdot\Vec{A},\label{London2}
\end{equation}
where $\Vec{q}$ is the wave vector normal to the semi-infinite superconductor surface, the second term on the right is called as the backflow. Here we consider the surface containing the vector potential and $x$-, $y$-axes [Fig.~\ref{setup}]. In this geometry, there is no backflow in the plane, thus the Eq.~(\ref{London2}) is written as
\begin{equation}
 \begin{pmatrix}
  j_{sx} \\
  j_{sy} 
 \end{pmatrix} 
 = -
 \begin{pmatrix}
  T_{xx} - \frac{T^2_{xz}}{T_{zz}} & T_{xy} - \frac{T_{xz}T_{yz}}{T_{zz}} \\
  T_{xy} - \frac{T_{xz}T_{yz}}{T_{zz}} & T_{yy} - \frac{T^2_{yz}}{T_{zz}} 
 \end{pmatrix}
 \begin{pmatrix}
  A_x(z) \\
  A_y(z)
 \end{pmatrix}.\label{London3}
\end{equation}

 In general, the symmetric tensor $\tilde{T}$ can be diagonalized for principle crystal axes, then we obtain the simple component form
\begin{equation}
-\frac{1}{\mu_0}\frac{\partial^2A_i}{\partial z^2} = j_{si} = -T_{ii}A_i = -\frac{1}{\mu_0\lambda^2_{ii}}A_i,
\end{equation}
for $i=x,y$, where $x,y,z$ axes are in parallel with principal crystal axes, we use Maxwell's equation $\mu_0\Vec{j}=\Vec{\nabla}\times(\Vec{\nabla}\times\Vec{A})$, $\mu_0$ is the permeability in vacuum, $\lambda_{ii}$ is the penetration depth defined by $A_i(z) = A_i(0)\exp{(-\lambda_{ii}z)}$. The normalized superfluid density is expressed as
\begin{equation}
    n_{ii}(T) = \left( \frac{\lambda_{ii}(T=0)}{\lambda_{ii}(T)} \right)^2 = \frac{T_{ii}(T)}{T_{ii}(T=0)}\label{nii}
\end{equation}

In our experimental configuration [Fig.~\ref{setup}], we set the $x$ and $z$-axes along [100] and [011] directions, respectively. Here $T_{xy}, T_{xz} = 0$, and we assume $q_z=0$ due to $\Vec{B}\parallel\Vec{z}$. Therefore Eq.(\ref{London3}) is written as
\begin{equation}
     \begin{pmatrix}
  j_{sx} \\
  j_{sy} 
 \end{pmatrix} 
 = -
 \begin{pmatrix}
  T_{xx}  & 0 \\
  0 & T_{yy}  
 \end{pmatrix}
 \begin{pmatrix}
  A_x \\
  A_y 
 \end{pmatrix}.
\end{equation}
From this equation, we obtain the penetration depth as $\lambda^2_{a}=\mu_0T_{aa}$. The symmetric tensor $\tilde{T}$ is expressed as
\begin{eqnarray}
    \tilde{T}(T) &=& \frac{e^2}{4\pi\varepsilon_0}\oint dS_F \left( \frac{\Vec{v}_F\Vec{v}_F}{v_F} \right)\left[ 1 - \frac{1}{2k_BT}\int_0^\infty d\xi_k \cosh^{-2}{\left( \frac{\sqrt{\xi^2_k + \Delta^2_k}}{2k_BT} \right)} \right],\label{T1}\\
    \tilde{T}(0) &=& \frac{e^2}{4\pi\varepsilon_0}\oint dS_F \left( \frac{\Vec{v}_F\Vec{v}_F}{v_F} \right),\label{T2}
\end{eqnarray}
where $dS_F$ is an element of Fermi surface area, and $\Vec{v}_F$ is the Fermi velocity. 


We use a simple model of anisotropic ellipsoidal Fermi surface that is defined by
\begin{equation}
E(\Vec{k})\frac{2m}{\hbar^2k_0^2} = \left(\frac{k_x}{A}\right)^2 + \left(\frac{k_y}{B}\right)^2 + \left(\frac{k_x}{C}\right)^2,
\end{equation}
where $E$ is the electron energy, $\hbar=h/2\pi$, $h$ is the Planck constant, $m$ is the electron mass, $k_0 = \sqrt{2mE_F}/\hbar$, $E_F$ is the Fermi energy, $A$, $B$ and $C$ are real numbers in the unit of \AA$^{-1}$. The Fermi wave vector $\Vec{k}_F$=$ k_F(\theta,\phi)$($\sin{\theta}\cos{\phi}$, $\sin{\theta}\sin{\phi}$, $\cos{\theta})$ is expressed in the polar coordinate as 
\begin{equation}
    k_F(\theta,\phi) = \left( \frac{\sin^2{\theta}\cos^2{\phi}}{A^2} + \frac{\sin^2{\theta}\sin^2{\phi}}{B^2} + \frac{\cos^2{\theta}}{C^2}\right)^{-1/2}.
\end{equation}
The Fermi velocity is expressed as, 
\begin{eqnarray}
    \Vec{v}_F &=& \frac{1}{\hbar}\Vec{\nabla}_k E(\Vec{k})|_{k=k_F}\\
    &=& \frac{E_F}{\hbar}\left(\frac{k_{Fx}}{A^2}, \frac{k_{Fy}}{B^2},\frac{k_{Fz}}{C^2}\right)\\
    v_F &=& \frac{E_F}{\hbar} k_F(\theta,\phi) \sqrt{\frac{\sin^2{\theta}\cos^2{\phi}}{A^4}+\frac{\sin^2{\theta}\sin^2{\phi}}{B^4}+\frac{\cos^2{\theta}}{C^4}}.
\end{eqnarray}
On this ellipsoidal Fermi surface model, the element of Fermi surface area is expressed as $dS_F = d\theta d\phi \sin{\theta}\sqrt{A^2B^2\cos^2{\theta}+C^2(B^2\cos^2{\phi}+A^2\sin^2{\phi})\sin^2{\theta}}$~\cite{Ellipsoid}. The ratio of anisotropic parameters  $A$, $B$, and $C$ are assumed from the experimental values of the directional dependence of the averaged Fermi velocity or the Fermi surface structure. Here we consider three cases: (I) Isotropic FS: The soft X-ray ARPES measurements observed the isotropic 3D Fermi surface~\cite{FujimoriJPSJ2019}. (II) Ellipsoidal FS (Fermi Surface): From the temperature gradient of the upper critical field, $(A,B,C)=(E_F/\hbar)$(1/11000 m/s, 1/5500 m/s, 1/9500 m/s)~\cite{AokiJPSJ2019}. (III) Open FS: The vacuum UV synchrotron ARPES measurements fitted the cylindrical light electron band with $(A,B,C)$=(0.313 \AA$^{-1}$, 1.103 \AA$^{-1}$, 0.729 \AA$^{-1}$) and the isotropic heavy electron pocket with $(A,B,C)$=(0.229 \AA$^{-1}$, 0.229 \AA$^{-1}$, 0.229 \AA$^{-1}$)~\cite{MiaoPRL2020}. The open light electron band is calculated by integrating the ellipsoidal model over the Brillouin zone ($|k_z|<0.45$ \AA$^{-1}$). In the cases I and II, the superfluid density only depends on the ratio of $A$, $B$, and $C$ because of the normalization of $T$ (see Eq.~\ref{nii}). We note that there are gaps in diagonal directions in measured Fermi surfaces. These gaps in the diagonal directions affect the integral over the Fermi surface, but in our model there is no any node near these directions. Thus here we consider this effect as small enough to be ignored.

In order to obtain $\Psi(T)$, we use the approximated form of 
\begin{equation}
    \frac{\Psi(T)}{k_BT_c}=\frac{\pi e^{-<\Omega^2\ln|\Omega|>}}{e^\gamma}\tanh\left( e^\gamma\sqrt{\frac{8(1-t)}{7\zeta(3)t}}\frac{e^{<\Omega^2\ln|\Omega|>}}{\sqrt{<\Omega^4>}} \right),
\end{equation}
where $<>$ is averaging over the Fermi surface, $<\Omega^2>=1$, $t=T/T_c$, $\zeta(3)\simeq1.2020$ is Riemann's zeta function, $\gamma\simeq0.577$ is the Euler constant~\cite{Kogan2021}. Because $\Psi(T)$ is obtained from the normalized $\Omega$, we only consider the ratio of $c_1, c_2$, and $c_3$.

From these results and Eqs.~(\ref{T1}) \& (\ref{T2}), we calculate the superfluid density $n_{(011)}\sim(n_a+n_{\perp[011],a})/2$ for possible gap symmetries [Figs. 4, S6, S7, S8 \& S9]. 
 
\begin{figure*}[tb]
\begin{center}
\includegraphics*[width=16cm]{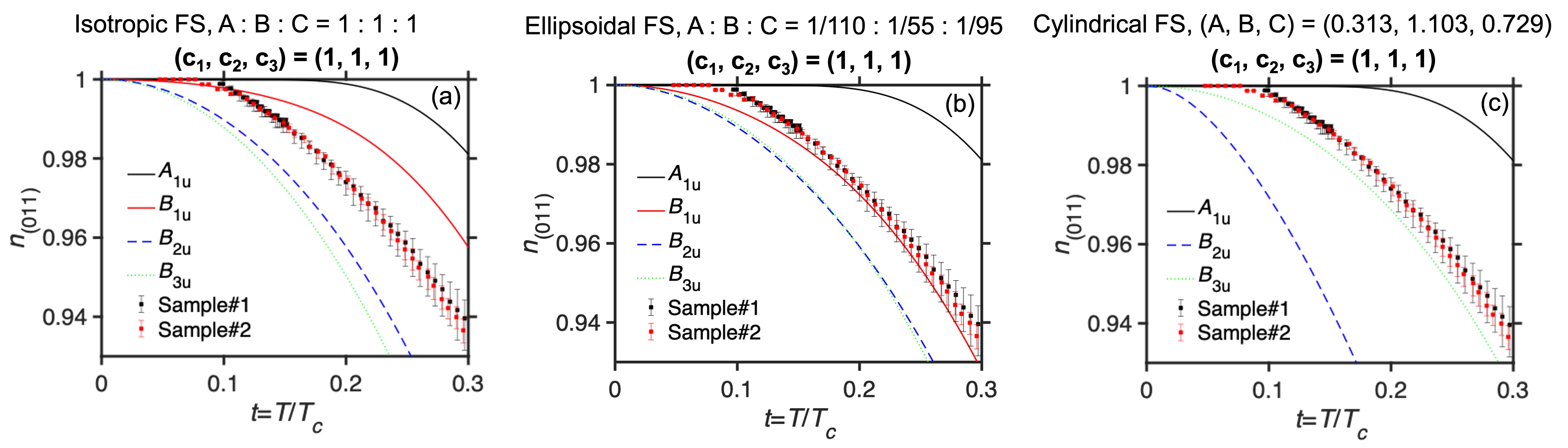}
\caption{Calculations of normalized superfluid density's temperature dependence with $c_1=c_2=c_3$ in (a) Isotropic FS case, (b) Ellipsoidal FS, (c) Cylindrical FS.}
\end{center}
\end{figure*}

\begin{figure*}[tb]
\begin{center}
\includegraphics*[width=16cm]{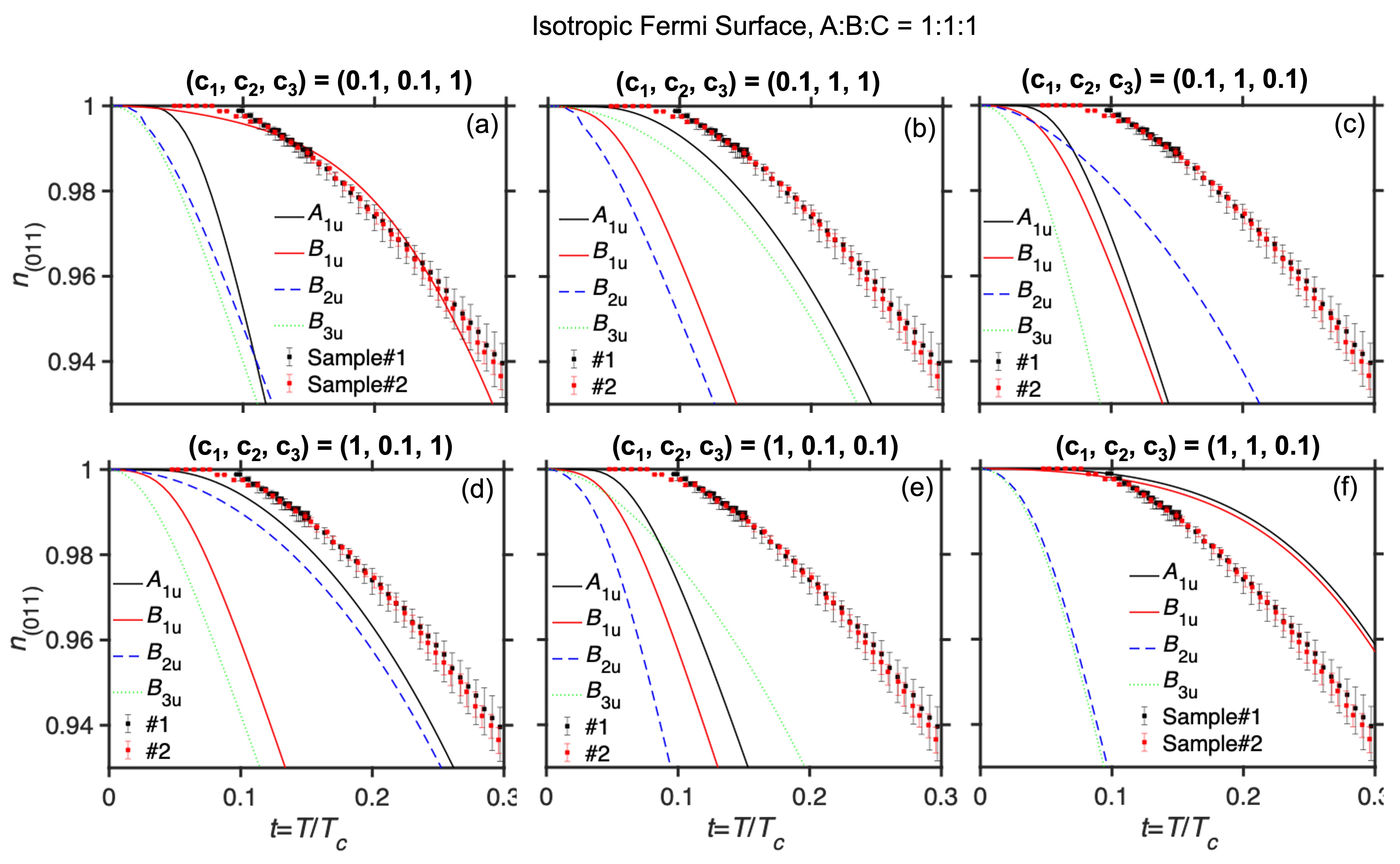}
\caption{Calculations of normalized superfluid density's temperature dependence in Isotropic Fermi surface case.}
\end{center}
\end{figure*}

\begin{figure*}[tb]
\begin{center}
\includegraphics*[width=16cm]{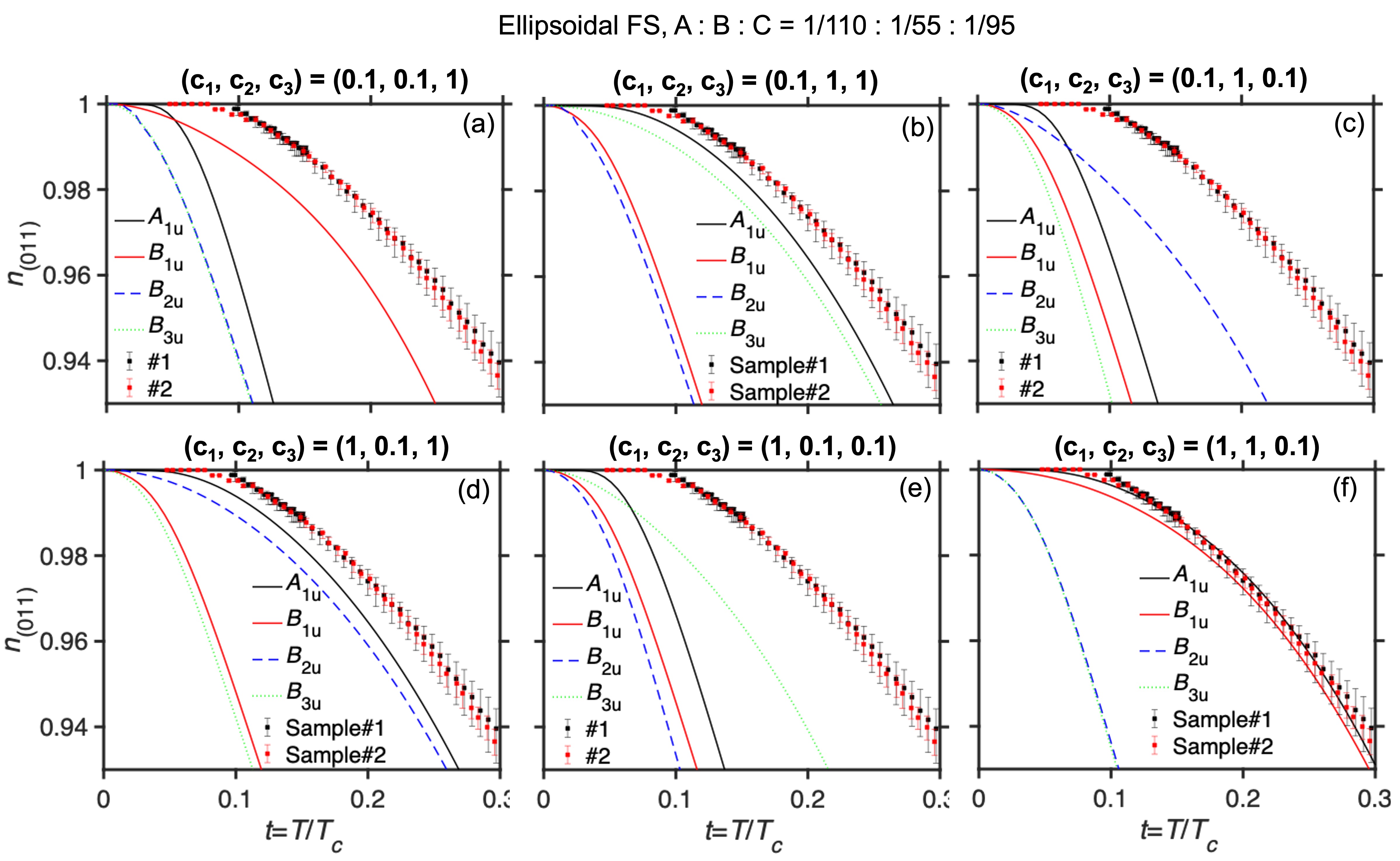}
\caption{Calculations of normalized superfluid density's temperature dependence in Ellipsoidal Fermi surface case.}
\end{center}
\end{figure*}

\begin{figure*}[tb]
\begin{center}
\includegraphics*[width=16cm]{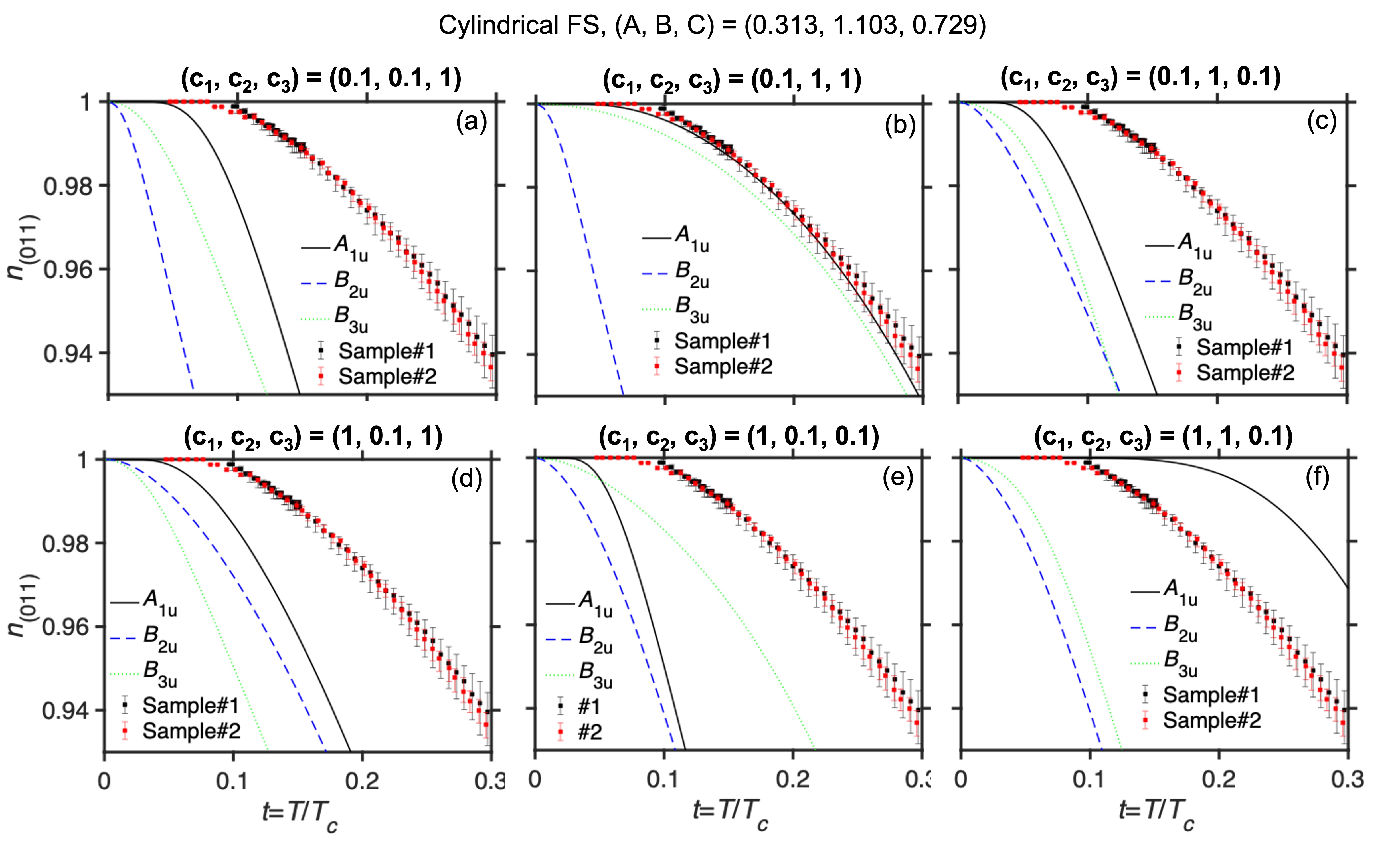}
\caption{Calculations of normalized superfluid density's temperature dependence in Cylindrical Fermi surface case.}
\end{center}
\end{figure*}

\end{document}